%% file: nad_revised.tex
\documentclass[referee]{emulateapj}
\usepackage{epsfig,graphicx,natbib}

\def\figspath{./}

\bibpunct{(}{)}{;}{a}{}{,}    %% A&A qfix to natbib
\def\cite#1{\citealp{#1}}     %% restore old astroncite \cite command
\input{rrmacros2e-AA}  %RR ?? add to final submission
\def\rmit#1{#1}               %% A&A & ApJ: latin abbreviations in Roman
\def\specchar#1{\uppercase{#1}}    %% RR for rrmacros2e-AA
\def\Caline{\CaII\ 8542~\AA} 

\usepackage{color}
\definecolor{lightred}{rgb}{0.8,0.0,0.0}
\definecolor{lighterred}{rgb}{0.8,0.45,0.45}
\definecolor{lightgreen}{rgb}{0.1,0.5,0.1}
\usepackage{ulem}

\begin{document} 

\title{The quiet solar atmosphere observed and simulated in \NaDone}

\author{J. Leenaarts$^{1,2,3}$}\email{jorritl@astro.uio.no} 
\author{R. J. Rutten$^{1,3}$}%\email{R.J.Rutten@uu.nl}
\author{K. Reardon$^{4,5,6}$}%\email{kreardon@arcetri.astro.it}
\author{M. Carlsson$^{1,2}$}%\email{mats.carlsson@astro.uio.no}
\author{V. Hansteen$^{1,2}$}%\email{viggo.hansteen@astro.uio.no}

\affil{$^1$ Institute of
  Theoretical Astrophysics, University of Oslo, P.O. Box 1029
  Blindern, N-0315 Oslo, Norway}
\affil{$^2$ Center of Mathematics for Applications,
  University of Oslo, P.O. Box 1053
  Blindern, N-0316 Oslo, Norway}
\affil{$^3$ Sterrekundig Instituut, Utrecht University, Postbus 80\,000,
         NL--3508 TA Utrecht, The Netherlands}
\affil{$^4$INAF -- Osservatorio Astrofisico di Arcetri,
         Largo Enrico Fermi 5, I-50125 Firenze, Italy}
\affil{$^5$ Astrophysics Research Centre, Queen's University, Belfast, BT7 1NN, Northern Ireland, UK}
\affil{$^6$ NSO/Sacramento Peak, P.O. Box 62, Sunspot, NM 88349--0062, USA}

\begin{abstract}
  The \NaDone\ line in the solar spectrum is sometimes attributed to
  the solar chromosphere.  We study its formation in quiet-Sun network
  and internetwork.  We first present high-resolution profile-resolved
  images taken in this line with the imaging spectrometer IBIS at the
  Dunn Solar Telescope and compare these to simultaneous chromospheric
  images taken in \Caline\ and \Halpha.  We then model \NaDone\
  formation by performing 3D NLTE profile synthesis for a snapshot
  from a 3D radiation-magnetohydrodynamics simulation.  We find that
  most \NaDone\ brightness is not chromospheric but samples the
  magnetic concentrations that make up the quiet-Sun network in the
  photosphere, well below the height where they merge into
  chromospheric canopies, with aureoles from 3D resonance scattering.
  The line core is sensitive to magneto-acoustic shocks in and near
  magnetic concentrations, where shocks occur deeper than elsewhere,
  and may provide evidence of heating sited deep within magnetic
  concentrations.
\end{abstract}


\section{Introduction}                              \label{sec:introduction}

%Studies of the fine-scale structuring of the lower atmosphere of the
%Sun advance the most when improved observation, in particular the
%higher angular resolution brought by adaptive optics and wavefront
%restoration, is combined with the improved understanding brought by
%increasingly realistic numerical simulation.  Photospheric quiet-Sun
%examples are the hydrodynamic formation of the solar granulation (see
%the review by
%
%\cite{2009LRSP....6....2N}) % Nordlund+Stein+Asplund review
%
%and reversed granulation
%
%(\cite{2005A&A...431..687L}, % Leenaarts+Wedemeyer tomo3
%\cite{2007A&A...461.1163C}), % Cheung++ reversed granulation
%
%the magnetoconvective origin of intergranular bright points marking
%kilogauss magnetic concentrations (\eg\
%
%\cite{2004ApJ...609L..91S}; %T Sanchez++, IN BPs
%\cite{2006A&A...452L..15L}; %C Leenaarts BP diagnostics
%\cite{2009A&A...499..301V}), % Vitas++ MnI
%
%and their appearance as faculae near the solar limb
%
%(\cite{2004ApJ...607L..59K}; % Keller++ BP simulation
%\cite{2004ApJ...610L.137C}). % Carlsson++ BP simulation limb
%
%These studies generally assumed local thermodynamical equilibrium
%(LTE) in both the numerical simulation and the subsequent
%spectral-line synthesis to emulate spectral diagnostics.

The past decade saw a rapid advance of our understanding of fine
structure and dynamics in the lower atmosphere of the Sun by
combination of high-resolution observations with increasingly
realistic numerical simulations. Examples are the formation of the
solar granulation
\citep{2009LRSP....6....2N} % Nordlund+Stein+Asplund review
and reversed granulation
(\cite{2005A&A...431..687L}, % Leenaarts+Wedemeyer tomo3
\cite{2007A&A...461.1163C}), % Cheung++ reversed granulation
and the magnetoconvective origin of intergranular bright points
\citep[\eg][]{2004A&A...427..335S}.
Such studies focused mainly on photospheric structure, and assumed
local thermodynamical equilibrium (LTE) in both the numerical
simulation and the subsequent spectral-line synthesis.

The present frontier in this type of research is to
study solar fine structure higher up in the solar atmosphere,
including the chromosphere. There the magnetic field fans out and becomes
more homogeneous, but shocks cause fine-scale structure and dynamics;
LTE is no longer valid.  Observationally, high-resolution and
rapid-cadence imaging is required in spectral lines that sample these
higher layers, with sufficient spectral resolution to chart and follow
their small-scale and fast profile variations. At present only imaging Fabry-P\'erot spectrometers at
ground-based telescopes with adaptive optics perform this task
adequately, sampling the few chromospheric lines in the red part of
the visible spectrum. 

In radiation-magnetohydrodynamics (MHD) simulations,
such upward advance requires full three-dimensionality and tackling
the challenging complexities of non-LTE (NLTE) radiative transfer and
non-equilibrium ionization/recombination balancing within the
simulation. Post-processing in the form of subsequent synthesis of
spectral diagnostics requires 3D NLTE radiative transfer, in some
cases even with partial frequency distribution.  In this paper we make
such an upward move for the quiet-Sun atmospheric regime sampled by
the the \NaDone\ resonance line at 5895.94~\AA.  This line is often
taken to be chromospheric, but we show here that it is much less so
than \Caline\ or \Halpha.

Recently, observational comparisons using Fabry-P\'erot spectrometry
were made between \CaII\ \HK\ and \Caline\ by
\citet{2009A&A...500.1239R} % Reardon++ H&K vs 8542
and between \Caline\ and \Halpha\ by
\citet{2009A&A...503..577C}. % Cauzzi++ caha
\citet{2009ApJ...694L.128L} % Leenaarts 8542 simulation
compared Fabry-P\'erot observations in \Caline\ with numerical
synthesis of this line from a 3D MHD simulation snapshot.  The present
paper is a companion to the latter study (hereinafter Paper~I), using
the same snapshot but turning to the \NaDone\ line.  The major
modeling advance in both papers is the incorporation of 3D NLTE
radiative transfer for the profile synthesis of the respective line.
Complete frequency redistribution is assumed and is in order
\citep{1989A&A...213..360U,1992A&A...265..268U}.

However, instantaneous LTE ionization equilibrium was assumed within
the MHD simulation itself. Most importantly, it was assumed
for hydrogen even though the studies by
\citet{2002ApJ...572..626C} % Carlsson+Stein dynamic H ionization
in 1D and by 
\citet{2007A&A...473..625L} % Leenaarts et al hion2
in 2D have shown that this is not an accurate assumption. Compared to
LTE, hydrogen is
underionized in hot shocks and overionized in cool post-shock
conditions. Both are due to a large ionization timescale
compared to the hydrodynamic evolution timescale. Improved treatment
is highly desirable, but its implementation in 3D MHD simulation codes
presents formidable challenges including severe processing-speed
requirements.  Paper~I and the present paper therefore represent steps
along the arduous obstacle course towards realistic {\it ab-initio}
simulation of the chromosphere, but not the final one.

In solar spectrum atlases \NaDone\ is about five times weaker than
\Caline\ in equivalent width, but its line center reaches deeper,
about 200~K lower in brightness temperature.  \NaDone\ and \Caline\
indeed differ in various formation properties. \NaDone\ is a strongly
scattering resonance line from an easily ionized minority stage;
\Caline\ is an excited line of the dominant calcium ionization stage
and has larger source-function coupling to the local temperature.
These differences make it worthwhile to employ and compare both lines.

In the next section we present and discuss high-resolution
spectrally-resolved imaging of a quiet-Sun area in \NaDone.  We then
use 3D NLTE synthesis of this line in the MHD simulation snapshot of
Paper~I to analyze its formation in quiet-Sun conditions
(Section~\ref{sec:simulations}).  We discuss the results in
Section~\ref{sec:discussion} and conclude the
paper in Section~\ref{sec:conclusions}.

\section{Observations} \label{sec:observations}

\subsection{Data acquisition}
%%%%%%%%%%%%%%%%%%%%%%%%%%%%%%%%%%%%%%%%%%%%%%%%%%%%%%%%%%%%%%%%%%%%%%%%%%%%

\begin{figure*}
  \includegraphics[width=\textwidth]{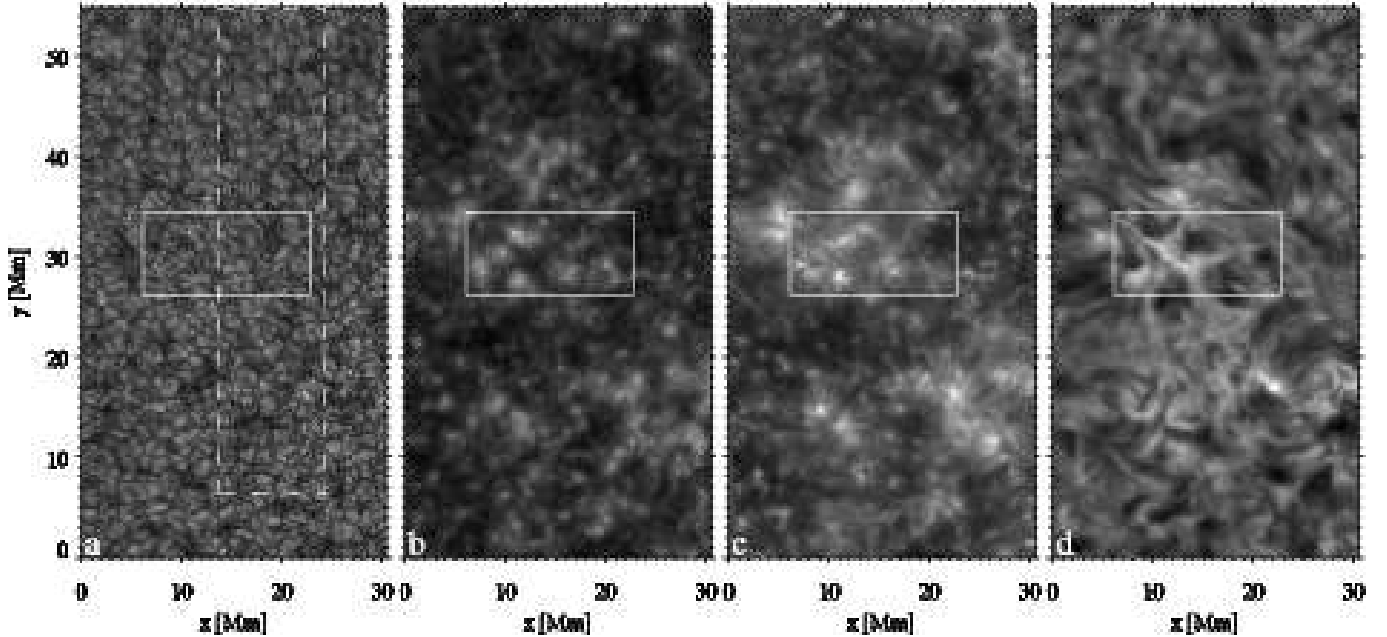}
  \caption{Images from the DST. Panels: (a) G~band, (b) \NaDone, (c)
    \Caline, (d) \Halpha. The latter three are constructed from the
    IBIS profile samplings by determining the intensity of the line
    profile minimum per pixel, in order to compensate for
    Dopplershifts from vertical bulk motions.  The dashed box shows
    the overlap with the Hinode/SP magnetograms in
    Figure~\ref{fig:obs-absB}.  The solid box is the subfield selected
    in Figure~\ref{fig:obs-center}.}
  \label{fig:obs-filter}
\end{figure*}

\begin{figure}
 \includegraphics[width=\columnwidth]{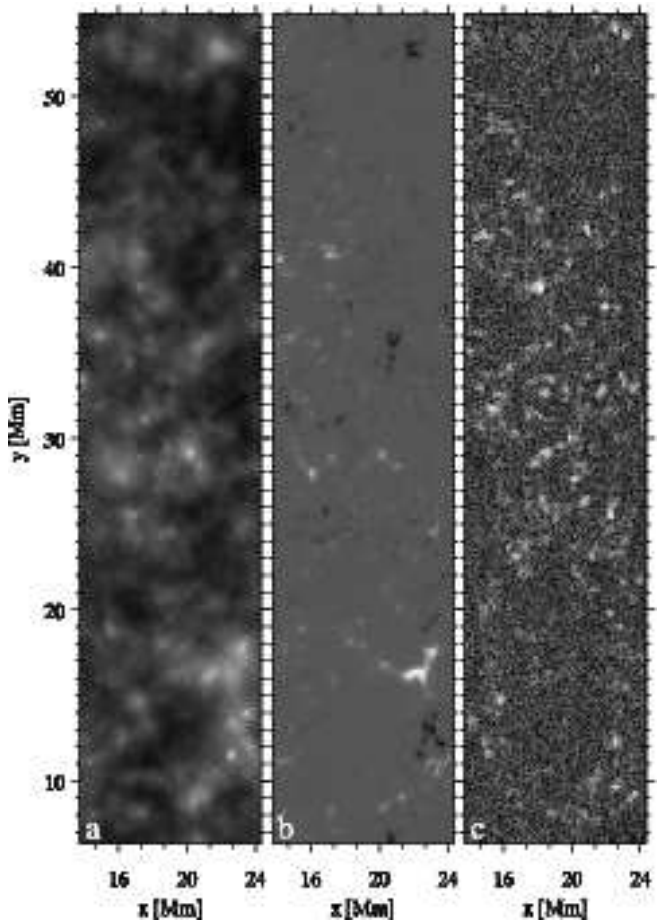}
 \caption{\NaDone\ profile-minimum intensity compared with
   longitudinal and transverse magnetograms for the overlap between
   the IBIS and Hinode/SP fields of view.  Panel (a): the intensity of
   the \NaDone\ profile minimum per pixel, as in
   Figure~\ref{fig:obs-filter}.  Panel (b): apparent longitudinal flux
   density.  Panel (c): apparent transverse flux density.  The flux
   density greyscale ranges from -441 to +629 Mx cm$^{-2}$ for panel
   (b) and 0 to +338 Mx cm$^{-2}$ for panel (c). The average measured
   flux density over the field-of-view is 70
   Mx~cm$^{-2}$. } \label{fig:obs-absB}
\end{figure}

\begin{figure*}
 \includegraphics[width=\textwidth]{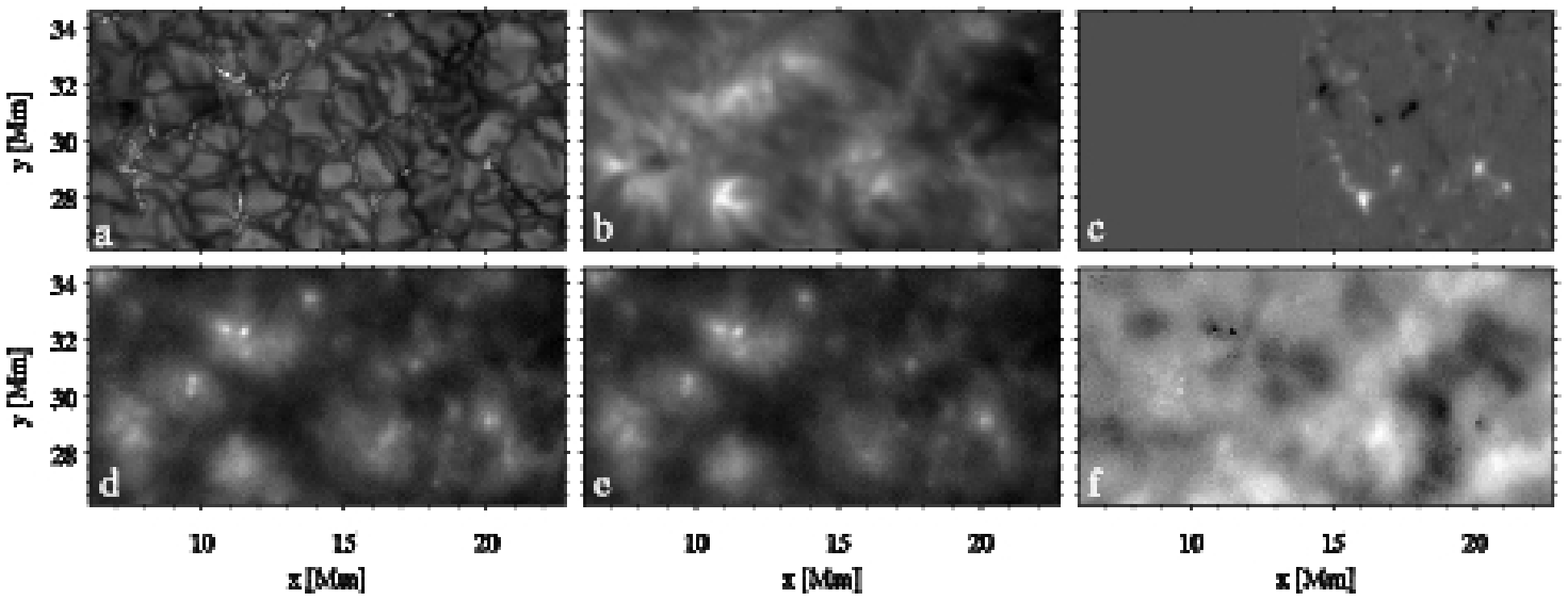}
 \caption{Details of the subfield marked by the solid box in
   Figure~\ref{fig:obs-filter}. Panels: (a) G~band, (b) \Caline, (c)
   longitudinal field, (d) \NaDone\ intensity at the nominal
   line-center wavelength, (e) \NaDone\ intensity of the profile
   minimum per pixel with the same greyscale as panel (d), (f)
   Dopplershift of the profile minimum per pixel, with the greyscale
   ranging from $-2$~\kms\ (dark, downflow) to $+2$~\kms\ (bright, upflow).}
\label{fig:obs-center}
\end{figure*}

We use imaging spectroscopy data obtained with the Interferometric
BIdimensional Spectrometer (IBIS) at the Dunn Solar Telescope (DST)
of the U.S. National Solar Observatory/Sacramento Peak.  IBIS consists
of two Fabry-P\'erot interferometers mounted in series
and delivers sequences of narrowband images with the spectral passband
stepping quickly through multiple user-selectable spectral lines
(\cite{2006SoPh..236..415C}; % Cavallini, IBIS I
\cite{2008A&A...481..897R}).  % Reardon+Cavallini, IBIS II
In this particular sequence, obtained during 15:30--17:07~UT on April
18, 2008, IBIS sampled \NaDone\ with a passband of 23~m\AA\ FWHM,
\Caline\ with a passband of 43~m\AA\ FWHM, \Halpha\ with a passband of
22~m\AA\ FWHM, and \FeI~7090~\AA\ with a passband of 24~m\AA\ FWHM.
Additional images were taken in the G~band with an independent camera
using an interference filter of 10~\AA\ FWHM.  One of the two
adaptive-optics feeds at the DST
(\cite{2004SPIE.5490...34R}) % Rimmele DST AO
served to stabilize the telescopic image and to correct wavefront
deformations in real time; the G-band images were additionally
improved through speckle reconstruction.  The $30\times54$~Mm$^2$
field of view covered a quiet solar area near the edge of an
equatorial coronal hole at apparent disk center (heliocentric longitude
0.56\deg~W, latitude 5.3\deg~S) with a pixel scale of 0.083
arcsec/pixel (close to critically sampling the 0.16 arcsec diffraction
limit of the 76 cm telescope at \NaDone). As part of a service-mode observing
program with Hinode a full-Stokes scan obtained with the Hinode/SP
spectropolarimeter in the \FeI\ 6302.5~\AA\ line between
14:32--16:00~UT overlapped with a portion of the IBIS field of view.
%\krb{I presume SOT/SP was actually sampling {\bf both} of the 6301/6302 Fe I lines?}

\subsection{Results}
%%%%%%%%%%%%%%%%%%%%%%%%%%%%%%%%%%%%%%%%%%%%%%%%%%%%%%%%%%%%%%%%%%%%%%%%%%%%
The IBIS images shown in
Figures~\ref{fig:obs-filter}--\ref{fig:obs-center} were taken between
15:54:53--15:55:17~UT and were selected as the sharpest of the sequence.  
The narrowband spectral images were additionally corrected by the application of destretching vectors
derived from simultaneous broadband reference images.
The dashed box in Figure~\ref{fig:obs-filter} specifies the overlap with
the Hinode/SP Stokes scan shown in Figure~\ref{fig:obs-absB}.  The
solid box outlines the subfield selected for
Figure~\ref{fig:obs-center}, chosen because it covers a similar area
as our simulation with a comparable amount of magnetic network,
including the most prominent bright points marking magnetic
concentrations.  However, only its right-hand half overlaps with the
Hinode/SP scan.

The four images in Figure~\ref{fig:obs-filter} differ clearly in their
portrayal of the solar atmosphere, ranging from the photospheric
G-band image at left to the chromospheric \Halpha\ image at right.
These two extremes display characteristic quiet-Sun scenes.  The
G-band image shows regular deep-photosphere granulation with relatively few
bright points marking strong-field magnetic concentrations, such as
the network patch within the solid box.  The \Halpha\ image shows
chromospheric fibrils (or alternatively ``mottles'', but we use the
term ``fibril'' for any elongated \Halpha\ structure), shorter and
more irregularly shaped than would be the case in regions with larger
magnetic activity. The fibrilar structure is more pronounced around
the photospheric network.

The two intermediate images differ significantly from both the outer
ones and also between each other.  The \Caline\ image shows some of
the same fibrils seen in \Halpha\ around denser network, but in the
quiet internetwork it shows a dark background with small roundish
brightness patches that are mostly due to acoustic shocks (\cf\
\cite{2009A&A...503..577C}). % Cauzzi++ caha
This regime was called ``clapotisphere'' by
\citet{1995ESASP.376a.151R}, % Rutten Asilomar review
denoting sub-canopy domains with sufficiently weak field that acoustic
and gravity waves can run up, become shocks, and interfere in
intricate patterns.  Similar patterns occur in field-free
hydrodynamical simulations
\citep{2004A&A...414.1121W}. % Wedemeyer++ CO5BOLD shocks

The \NaDone\ image (second panel) shows essentially no fibrilar structure,
with only occasional glimpses of short and bright elongated
structures. The brightest G-band bright points show up as co-spatial
bright-point features, but with wide bright aureoles around them.  The
dark internetwork areas show vague pattern similarities with the
\Caline\ image.  Thus, at first sight \NaDone\ seems to sample the
same internetwork waves, and possibly clapotispheric shocks, with
photospheric magnetic concentrations visible as bright points
with aureoles. The same is shown in the similar but
lower-resolution images of
\citet{2002A&A...383..283A}.

With the lack of fibrils and only a few strong bright points in this quiet-Sun region,
the \NaDone\ image appears rather featureless. This apparent blandness
is a solar property, since
the broadband images taken simultaneously with the spectral images show excellent 
sharpness, especially near the adaptive-optics lock point (located 
in the subfield shown in Figure~\ref{fig:obs-center}). 
\citet{righini++2009submitted} %Righini et al. (2009)
have shown that 
IBIS achieves near diffraction-limited image quality in its narrowband channel.
Only small-scale features with very low contrast might drown in the photon
noise set by the average line-core intensity of
2000~photons/pixel during the 50~ms exposures. 
The variation in the appearance of the bright aureoles among 
different bright points and at different wavelengths shows that they
do not result from scattering in the instrument or the Earth's atmosphere.

Figure~\ref{fig:obs-absB} confirms the identification of \NaDone\
bright points as magnetic concentrations.  This part of the Hinode/SP Stokes
scan lasted from 15:51:23 until 15:59:38~UT, so that it is not fully
co-temporal with the \NaDone\ images. Nevertheless, there is
remarkable correspondence between the brightest \NaDone\ features and
strong longitudinal field, much more than with the (generally much
weaker) transversal field.

Figure~\ref{fig:obs-center} contains enlargements of the selected
subfield.  The magnetogram in panel (c) shows the signed longitudinal
flux density but has only partial overlap.  Panel (d) adds a
filtergram display of the \NaDone\ intensity, \ie\ with the IBIS
passband at the nominal rest wavelength of the line, as compared to
the image generated using local Doppler compensation in panel (e).
The two panels have the same greyscale.  The brightest features are
brighter in panel (d) due to large central downflows which can be seen
in panel (f) as three tiny very dark patches in the middle of magnetic
concentrations.  Three other magnetic concentrations display
comparably narrow updrafts.  These motions Doppler-shift the line core
out of the passband, causing additional brightening, as also pointed
out by
\citet{2002A&A...383..283A}. %Al,Bendlin & Kneer NaD2
Such mixing of Dopplershift and intensity modulation generally affects
standard fixed-wavelength narrow-band filtergrams. We therefore plot
Doppler-compensated line-center intensities rather than
rest-wavelength line-center intensities, so that brightness variations
are more indicative of the source function than of the velocity field
in the line forming region.
%It validates our
%approach of plotting Doppler-compensated line-center intensities
%rather than rest-wavelength line-center intensities, exploiting the
%profile sampling capabilities of IBIS.
The larger-scale patterning in panel (f) has no obvious correspondence
to the other panels.

%The upshot of this section is that the \NaD1\ image looks much more
%like an unsigned photospheric Stokes-$V$ magnetogram than it looks
%chromospheric. 
The upshot of this section is that the \NaD1\ core brightness seems to
coincide more with the photospheric Stokes-$V$ magnetogram signal than
with fibrilar structure seen in \Caline\ and \Halpha.  This suggestion
agrees with the earlier conclusion of
\citet{2000A&A...357.1093C} % Cauzzi++ NaD vs HK
from lower-resolution (spatial and spectral) data that the \NaDtwo\
core intensity correlates well with the apparent magnetic flux density
in network (above 50~Mx~cm$^{-2}$ in their Figure A1).
Figure~\ref{fig:obs-filter} indicates that \NaDone\ in the
internetwork may show some clapotispheric intensity modulation caused
by waves and shocks, mimicking those in \Caline, but none of the
chromospheric structure seen in \Halpha.

%%%%%%%%%%%%%%%%%%%%%%%%%%%%%%%%%%%%%%%%%%%%%%%%%%%%%%%%%%%%%%%%%%%%%%%%%%%%
\section{Simulations} \label{sec:simulations}
%%%%%%%%%%%%%%%%%%%%%%%%%%%%%%%%%%%%%%%%%%%%%%%%%%%%%%%%%%%%%%%%%%%%%%%%%%%%

\subsection{MHD simulation snapshot}
%%%%%%%%%%%%%%%%%%%%%%%%%%%%%%%%%%%%%%%%%%%%%%%%%%%%%%%%%%%%%%%%%%%%%%%%%%%%
We used a snapshot from a radiation-MHD simulation computed with the
Oslo Stagger Code (OSC, 
\cite{2007ASPC..368..107H}).  % Hansteen++ Coimbra
The same snapshot was used in Paper~I.  OSC includes an LTE
equation of state based on the Uppsala Opacity Package
\citep{gustafsson1973}
and uses radiative transfer with multi-group opacity binning following
\citet{1982A&A...107....1N} % Nordlund: solar granulation I
and NLTE scattering following
\citet{2000ApJ...536..465S} % Skartlien: multigroup scattering.
to compute radiative losses in the photosphere and low chromosphere.
Radiative losses in the middle and upper chromosphere are evaluated
with an escape probability method including continua and lines of
hydrogen plus lines of \CaII.  It was calibrated with the 1D RADYN
code
\citep[\eg][]{1997ApJ...481..500C}. %C C+S H2v grain simulation
Higher up, OSC employs optically thin radiative cooling.  Thermal
conduction along magnetic field lines is taken into account. More
detail is given in
\citet{2007ASPC..368..107H} % Hansteen++ Coimbra
and Mart\'{\i}nez-Sykora \etal\ (2008, 2009).
\nocite{2008ApJ...679..871M} % Martinez-Sykora et al: flux tube emergence
\nocite{2009ApJ...701.1569M} % Martinez-Sykora et al: spicule simulations

The simulation used here has a grid of $256 \times 128 \times 160$,
corresponding to a physical size of $16.6 \times 8.3 \times
15.5$~Mm$^3$. It extends from 1.5~Mm below the photosphere to 14~Mm
above it, \ie\ from the upper convection zone into the corona. The
grid has 65~km horizontal resolution. The grid is non-uniform in the
vertical direction; its resolution varies from 32~km in the convection
zone to 440~km in the corona.  The snapshot contains bipolar magnetic
%MJ would prefer gauss as unit for field strength in simulation
field with a mean strength of 150~gauss in the photosphere.

\subsection{Radiative transfer}

\begin{figure}
  \includegraphics[width=\columnwidth]{\figspath/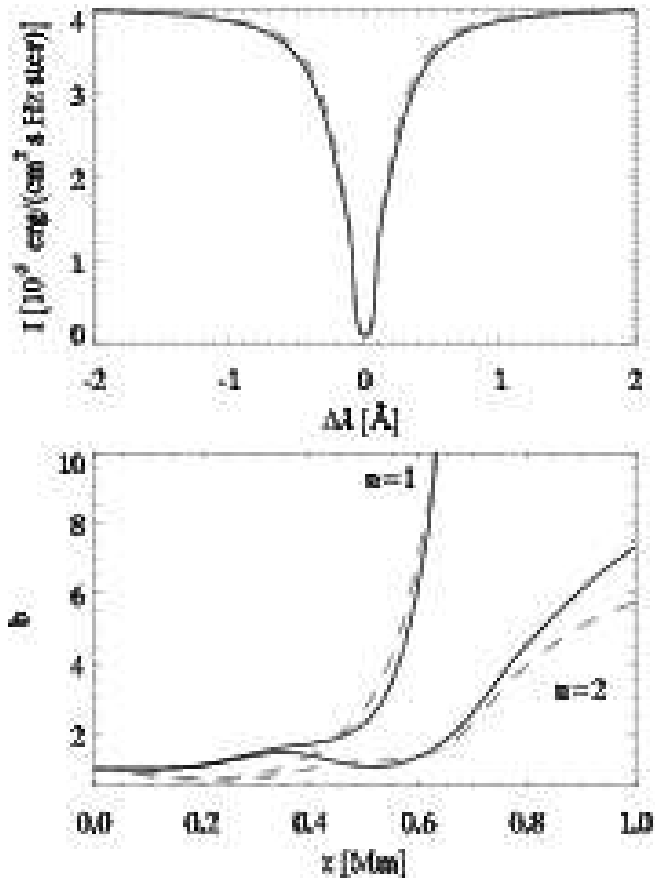}
  \caption{Comparison of \NaDone\ line formation in the FALC
    atmosphere between the 18-level (black curves) and 4-level (grey
    dashed curves) model atoms. Upper panel: emergent line profiles at
    $\mu \is 0.95$.  Lower panel: NLTE population departure
    coefficients for the lower ($n \is 1$) and upper ($n \is 2$) level
    of the line.}
\label{fig:atomtest}
\end{figure}

For the radiative transfer computation of the emergent \NaDone\
intensity the atmosphere was interpolated in the vertical direction
onto a grid with 153~points covering a height range between $-0.5$ and
2~Mm relative to continuum optical depth $\tau_{5000} \is 1$, covering
the formation range of all pertinent sodium transitions. The electron
density was computed assuming LTE ionization for all relevant species,
including hydrogen, consistent with the LTE equation-of-state used in
OSC. The angle quadrature was computed using the A4 set of
\citet{Carlson1963}, %Methods of Computational Physics, Vol.~1 TEMP
employing 3 rays per octant.  The gas motions in the simulation were
taken into account without adding additional microturbulence.

The radiative transfer computation was performed using MULTI3D
\citep{botnen1997,leenaarts+carlsson2009}, % Hinode 2 conf procs
a code based on the 1D code MULTI of
\citet{1986UppOR..33.....C} % MULTI Uppsala report
and including the same physics, but with a 3D short-characteristics
solver allowing evaluation of the 3D radiation field.  MULTI3D has
been MPI-parallellized using domain decomposition so that it runs
efficiently on supercomputers.

Since 3D radiative transfer remains very demanding, the actual \NaI\
Grotrian diagram was simplified to a minimal model atom.
Inspired by the results of
\citet{1992A&A...265..237B}, % Bruls++ alkali I
we constructed a model atom of only four levels, consisting of the
ground state, the upper level of \NaDone, an extra level at 1~eV below
the continuum, and the \NaII\ ground state as continuum.  We gave the
extra level normal collisional coupling to the two lower levels but
enormous ($10^5$ in excess of normal values) collisional bound-free
coupling to the continuum and large (Gaunt factor 2.2) radiative
coupling to the upper level of \NaDone.  We added this extra level to
implement the suction process identified by
\citet{1992A&A...265..237B} % Bruls++ alkali I
which draws population from the ion reservoir into the \NaD\ ground
state. We adjusted its strong couplings such that it has similar suction
power as the complete full-atom upper term structure, which we
determined by comparing the resulting \NaDone\ level populations to
those of the full 18-level model atom specified in Table~2 of
\citet{1992A&A...265..237B} % Bruls++ alkali I
for the standard 1D FALC atmosphere of
\citet{1993ApJ...406..319F}. % Fontenla et al.
The result is shown in Figure~\ref{fig:atomtest}.  The emergent
profiles in the upper panel are nearly equal.  The lower panel
compares the NLTE population departure coefficients $b_\mathrm{i}
\equiv n_\mathrm{i}^\mathrm{NLTE} / n_\mathrm{i}^\mathrm{LTE}$ for the
lower and upper \NaDone\ level between the two model atoms.  They
match very well, especially over the most important range $z \is
0.4-0.6$~Mm where the line core becomes optically thin.  The steep
increase of the $b_1$ departure coefficient is due to the dominance of
photoionization over collisional ionization, but sufficient suction is
needed to match it with the small atom.  Slightly larger differences
occur deeper down, where both small-atom curves miss the slight
overpopulation in the FALC temperature minimum due to lack of photon
suction in weaker lines. However, the ratio $b_2/b_1$ remains nearly
equal between the two models so that the \NaDone\ source functions are
nearly equal; only the optical-depth buildup below the formation
height of the line core differs somewhat.  This simple model atom is
accurate enough for our purposes.  We used it to compute emergent
\NaDone\ profiles from the simulation snapshot, both treating each
column as a plan-parallel atmosphere and in full 3D, assuming complete
redistribution which is a good assumption for this line
\citep{1992A&A...265..268U}. % Uitenbroek & Bruls: PRD in NaI and BaII lines

\subsection{Snapshot properties}

\begin{figure*}
  \includegraphics[width=\textwidth]{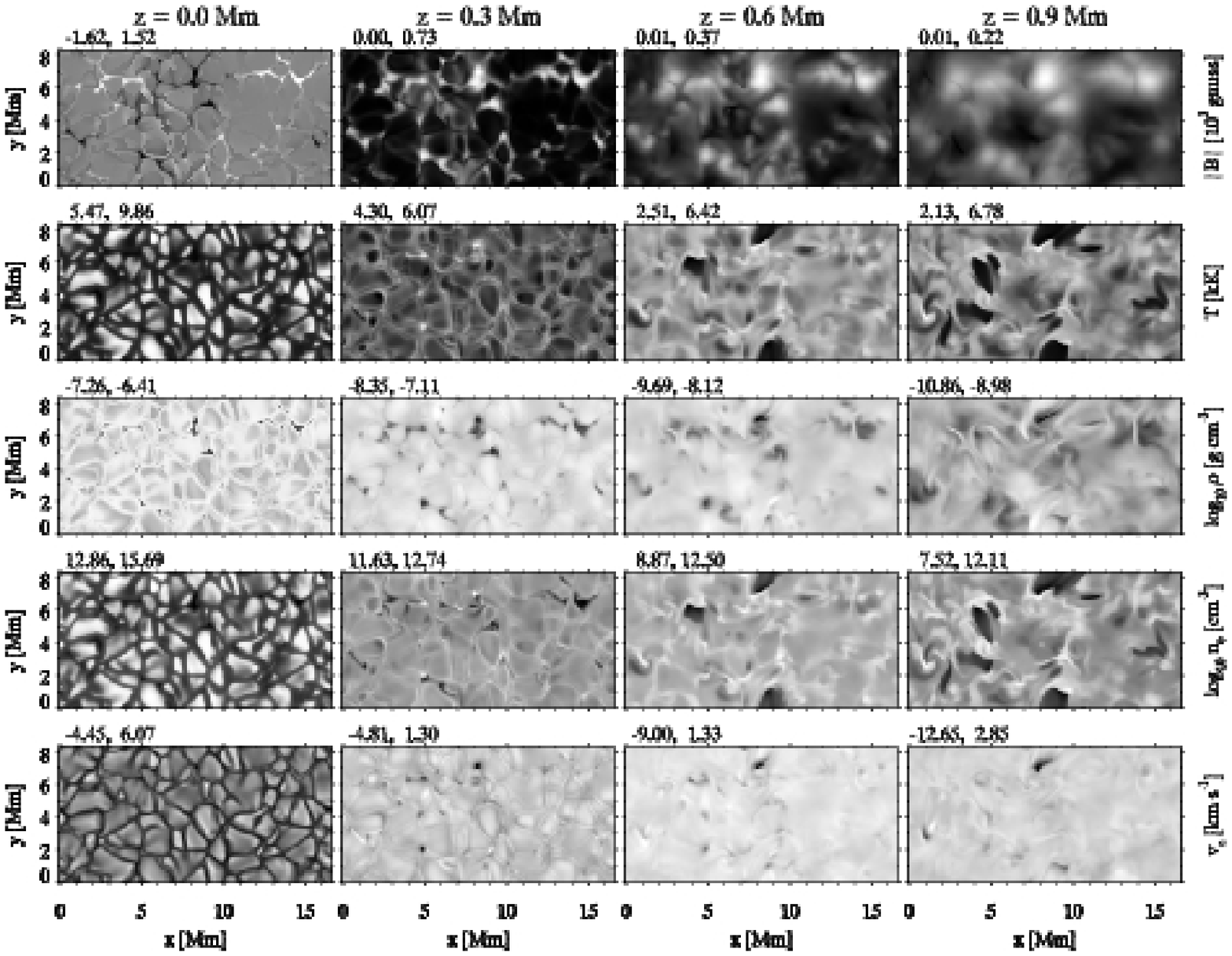}
  \caption{Properties of the simulation snapshot. All panels show
    $xy$-cuts through the simulation cube at the heights specified at
    the column tops, with $z \is 0$~Mm corresponding to
    spatially-averaged $\tau_{5000} \is 1$.  Each row displays the
    quantity specified with its units at the right-hand side, from top
    to bottom: magnetic field strength, gas temperature, mass
    density, electron density, vertical velocity (bright for upflow).
    Each panel is greyscaled to maximum contrast, with the range
    specified along each panel top. The top-left panel displays the
    magnetic field strength multiplied with the sign of
    $B_\mathrm{z}$ (upwards bright) to show the polarity
    distribution.}
  \label{fig:sim-MHD}
\end{figure*}

Figure~\ref{fig:sim-MHD} displays simulation results in the form of
horizontal cuts through the snapshot cube at four different heights,
specified at the top of each column.

The first row shows how the magnetic field is organized in small-scale
intergranular magnetic concentrations, especially at lane vertices,
that expand with height as in the classical fluted fluxtube paradigm
(\eg\ \cite{1977PhDT.......237S}; %B thesis
Solanki 1987, 1993).  \nocite{1987PhDT.......251S} % Solanki thesis
\nocite{1993SSRv...63....1S} %C Solanki new testament
At $z \is 0.0$~Mm (upper-left panel) they do not appear as roundish
fluxtubes but rather as extended features similar to the G-band
``ribbons'' and ``flowers'' in observations with comparable resolution
\citep{2004A&A...428..613B,2005A&A...435..327R}. % Rouppe+ flower, ribbon evolution
The expansion with height is striking (note the scale reduction to
seven times lower amplitude at $z \is 0.9$~Mm).  The actual field is
bipolar with a roughly even split between positive and negative
concentrations, as shown in the first panel (where the field is mostly
near-vertical).

The temperature cut at $z \is 0.0$~Mm in the second row shows
granulation. At $z \is 0.3$~Mm the brightness pattern has changed into
reversed granulation; the strongest magnetic concentrations stand out
by being hotter than their surroundings at equal geometrical height.
Higher up, fairly large dark patches (``clouds'') of very cool gas
occur, with hot white borders indicative of shocks.  They occur
preferentially near magnetic polarity changes.

The gas density (third row) shows the familiar evacuation of magnetic
concentrations.  At $z \is 0.0$~Mm these are superimposed on the
higher-density intergranular lanes that appear as negative of the
temperature pattern.  At $z \is 0.3$~Mm the tops of the largest
granules are densest.  The magnetic evacuation persists at all sampled
heights, spreading with the field strength but maintaining
co-spatiality with the photospheric topology up to $z \is 0.6$~Mm.  In
the $z \is 0.9$~Mm panel the gas density shows slender high-density
features that likely mark shock interference.

The electron density (fourth row) mimics the temperature closely at
all heights because both the hydrogen ionization balance and the
electron-donor (Si, Fe, Al, Mg, Ca, Na) ionization balances are
computed in LTE.  At $z \is 0.3$~Mm most magnetic concentrations have low
electron density due to low overall gas density, but the hottest show
excess electron density from donor ionization.  The $N_\rme/N_\rmH$
ratio of electron to hydrogen density (not shown) reaches values
as low as 10$^{-6}$ in the coolest clouds in the higher layers, well
below the $10^{-4}$ relative abundance of the electron donor elements,
implying that even these are largely neutral.  The thin
high-temperature shocks around the cool clouds have large electron
densities.

The vertical velocity pattern in the bottom row shows the familiar
pattern of granular upflow and downflow in the photosphere.  Its
amplitude diminishes rapidly with height.  Higher up, the most
striking features are large downdrafts in most of the stronger
magnetic concentrations, in particular the strongest one at $(x,y) \is
(8.2,7.2)$.  They survive up to $z \is 0.9$~Mm.  The upflow patterns
resemble the gas density patterns.

\subsection{Emergent \NaDone\ profiles}
\label{sec:emergent}

\begin{figure}
  \includegraphics[width=\columnwidth]{\figspath/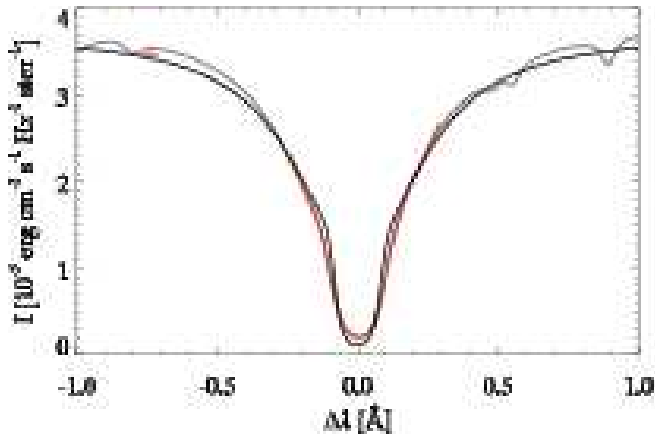}
  \caption{Comparison of spatially-averaged observed and simulated
    \NaDone\ line profiles. Grey curve: averaged disk-center intensity
    from the NSO Fourier Transform Spectrometer (FTS) atlas calibrated
    by \citet{1984SoPh...90..205N}. % Neckel+Labs
    Red curve and red plus sign: IBIS observation.  Black curve:
    simulation. Both the FTS atlas and the simulation provide absolute
    intensities. The IBIS data were scaled to match the FTS profile in
    the line core.}  %\krb{Since the plot already requires color to be legible, maybe
    %make the simulated profile another color?}}
  \label{fig:sim-profiles}
\end{figure}

Figure~\ref{fig:sim-profiles} compares the spatially-averaged \NaDone\
profile from the simulation with the spatially-averaged IBIS
observation and with a solar-atlas profile. The agreement between the
IBIS and atlas  profiles is excellent. The simulation deviates in
producing a slightly deeper core and narrower inner wings, with more
pronounced knees. This lack of core broadening is similar to, but
smaller than for the \Caline\ line in Figure~1 of Paper~I.  The
latter was attributed, at least partially, to a lack of spatial
resolution in the microturbulence-free simulation.  Since we use the
same snapshot, this deficiency also applies here. However, its effect
is probably smaller because \NaDone\ is formed lower than \Caline\
whereas the temperature and density in Figure~\ref{fig:sim-MHD} show
an increase in fine-structure amplitude with height.

\begin{figure}
\includegraphics[width=\columnwidth]{\figspath/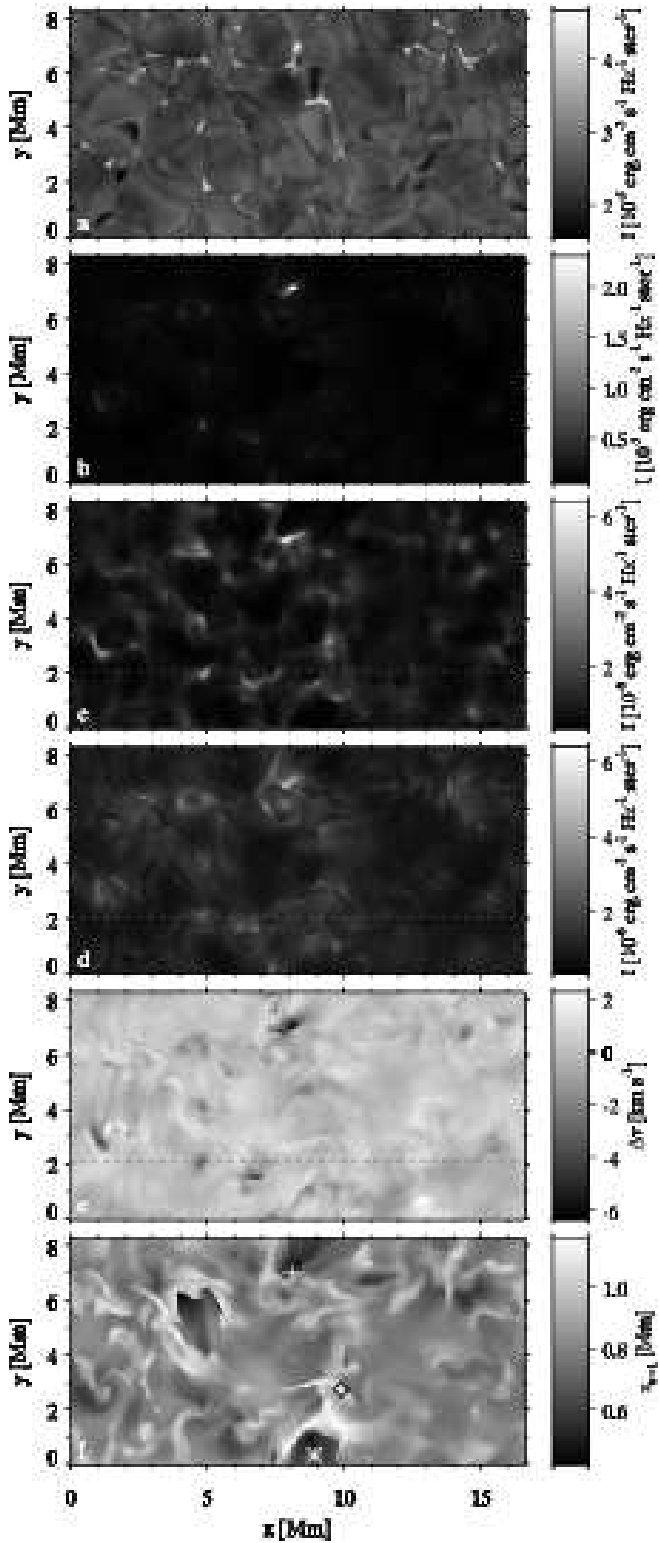}
\caption{\NaDone\ synthesis with the 3D solver, except for panel (c).
  The intensity maps show the vertically emergent intensity from the
  simulation.  From top to bottom: (a) intensity in the blue wing at
  $\Delta \lambda \is -0.27$~\AA\ from nominal line center; (b)
  intensity at the nominal line center (rest wavelength); (c)
  intensity in the profile minimum per pixel, using the 1D solver; (d)
  intensity in the profile minimum per pixel; (e) Dopplershift of the
  profile minimum per pixel (upflows bright); (f) $\tau \is 1$ height
  for the profile minimum per pixel.  The dotted line near $y \is
  2.0$~Mm specifies the cut location used in
  Figure~\ref{fig:sim-3D-1D}.  The three symbols in the bottom panel
  specify the locations for which \NaDone\ formation breakdown is
  given in
  Figure~\ref{fig:sim-breakdown}.}
\label{fig:sim-images}
\end{figure}

Figure~\ref{fig:sim-images} displays properties of the \NaDone\
spectral profiles that emerge vertically from the top of the simulation volume,
in the form of  images comparable to the lower row in
Figure~\ref{fig:obs-center}.  The wing image in the first panel serves
as low-photosphere indicator, showing the magnetic concentrations as bright
points.  The blue \NaDone\ wing was not included in our earlier survey
of proxy-magnetometry diagnostics
(\cite{2006A&A...452L..15L}), %C Leenaarts BP diagnostics
but it is comparable to the other strong-line wings there. The
magnetic concentrations gain contrast because of weaker damping wings,
and, in the case of \NaDone\ also because of enhanced ionization from
partial evacuation.

The second panel of Figure~\ref{fig:sim-images} emulates a narrow-band
filtergram with the passband at nominal line center.  The strongest
magnetic concentration appears very bright, mostly through downdraft
Dopplershift as demonstrated in Figure~\ref{fig:sim-breakdown}.  The
rest of the scene vanishes by setting the greyscale to include this
feature.

Panels (c) and (d) compare 1D plan-parallel per-column radiative
transfer with full 3D radiative transfer, plotting the profile-minimum
intensity per pixel in order to compensate for bulk-motion
Dopplershifts.  The greyscales are identical.  The scenes are
comparable but the magnetic concentrations reach larger contrast in
the 1D case.  In 3D, radiation scatters out of these and illuminates
their surroundings, as shown below.

Panel (e) maps the Dopplershift of the line profile minimum per
pixel. The greyish small-amplitude bright and dark patterning
resembles the $z \is 0.9$~Mm vertical-velocity sampling in the last
panel of Figure~\ref{fig:sim-MHD}.  The larger magnetic-concentration
downdrafts mimic those in the $z \is 0.6$~Mm panel most closely.
Narrow bright structures due to upward-propagating shocks appear
elsewhere.

Finely structured bright shocks and dark clouds appear abundantly in
panel (f), which displays the geometrical height where radiation at
the wavelength of the profile minimum per pixel reaches optical depth
unity.  The range is very large, varying from $z \is 0.5$~Mm in the
cool clouds to $z \is 1.1$~Mm in the shock fronts.  This rugged
height-of-formation pattern is effectively a map of the \NaI/\NaII\
ionization-recombination balance and is closely similar to the
electron density pattern at $z \is 0.9$~Mm in
Figure~\ref{fig:sim-MHD}, adding the slender high-density features in
the panel above it. 
%\krb{This is a bit confusing. If $h_\tau=1$ varies so much why would it resembles
%$n_e$ at z=0.9 Mm?}

In the cool clouds sodium is nearly completely ionized. This
counterintuitive result (one would expect sodium to be largely neutral
at these temperatures and densities) is explained as follows: The
electron density in the model atmosphere is computed in LTE and the
cool clouds have thus a low electron density. The NLTE sodium
ionization-recombination balance is set by the interplay of
photoionization versus photorecombination requiring one electron.  The
photoionization rate per sodium atom is independent of temperature, as
it is set by the radiation field. The recombination rate per sodium
atom however depends on the electron density. In the low-temperature
clouds the electron density is low, thus the recombination rate is low
and as a consequence the ionization-recombination balance shifts
towards \NaII.

The \NaDone\ $\tau\is1$ height map obtains its enormous variation and
fine structure from the large variations in electron density affecting
the sodium recombination.  The Dopplershift map in panel (e) follows
this $\tau\is1$ height variation; it samples the Dopplershift in
magnetic concentrations and cool clouds deeper in the atmosphere than
in hot shocks..

Note that the $\tau\is1$ height variation would be even larger if we
had not plotted $\tau_\nu \is 1$ for the Doppler-compensated profile
minimum per pixel but $\tau_{\nu_0} \is 1$ at the nominal line center
as in a narrow-band filtergram.  In that case Dopplershifts would
cause even deeper line-center formation, as deep as $z \is 0.3$~Mm in
the downdrafts in the magnetic concentrations
(Figure~\ref{fig:sim-breakdown}).

\subsection{Comparison of 3D and 1D modeling}

\begin{figure*}
\includegraphics[width=\textwidth]{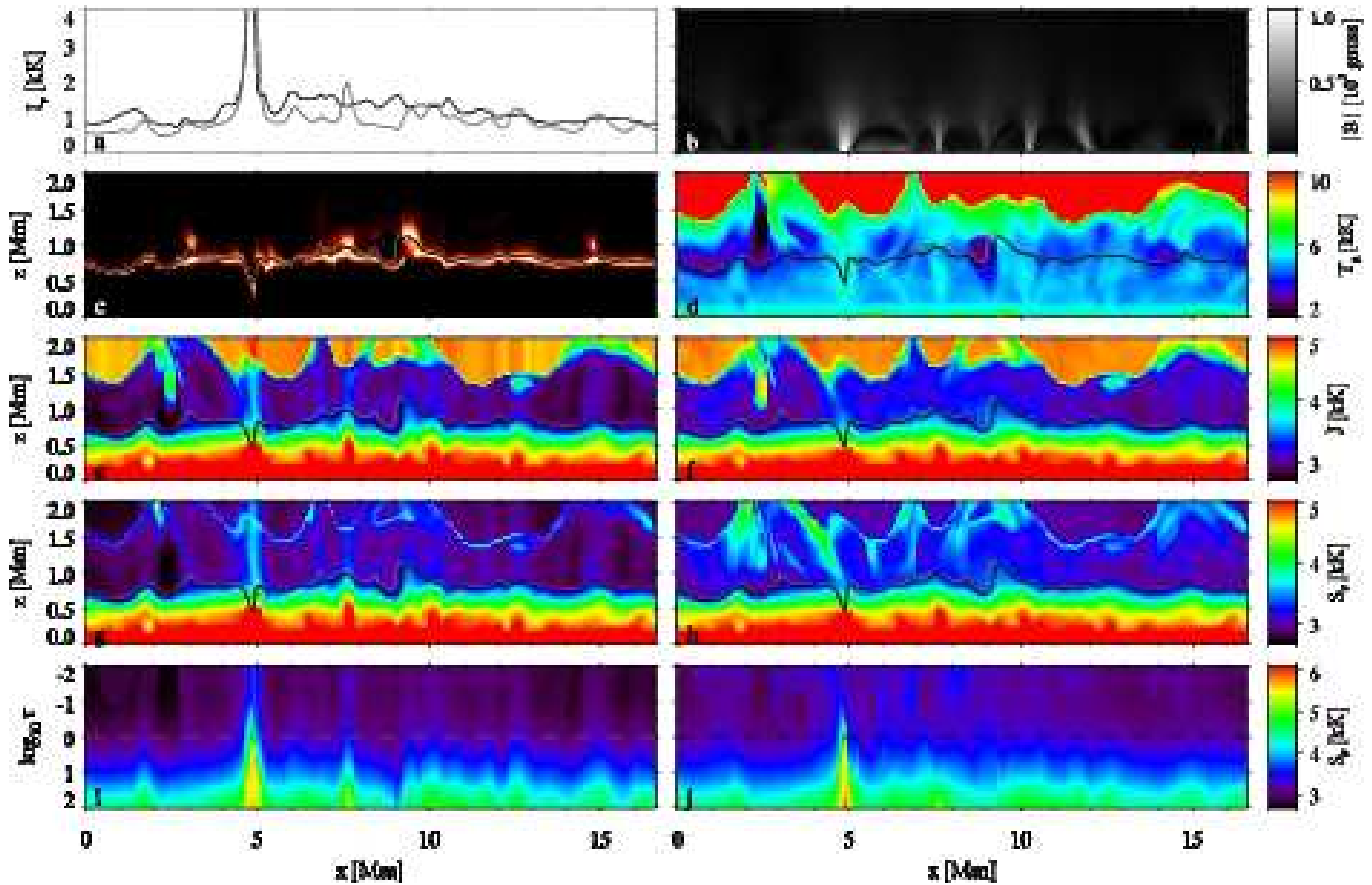}
  \caption{Comparison between 1D and 3D \NaDone\ line-center formation
    along the $xz$-cut near $y \is 2.0$~Mm marked by the dotted line in
    Figure~\ref{fig:sim-images}.  Panels (c) to (h) have the $\tau\is$
    height overplotted for the 3D (black) and the 1D solver (grey).
    Panels: (a) emergent intensity for the 3D (black curve) and 1D
    (grey curve) solver in radiation temperature units; 
    (b) magnetic field strength; 
    (c) contribution function for the 3D solver; 
    (d) gas temperature; (e) 1D radiation field; (f) 3D radiation
    field; (g) 1D source function; (h) 3D source function; (i) 1D
    source function on line-center $\tau_{\nu_0}$-scales per column;
    (j) 3D source function on line-center $\tau_{\nu_0}$-scales per
    column.}
  \label{fig:sim-3D-1D}
\end{figure*}

Figure~\ref{fig:sim-3D-1D} compares our 3D evaluation of \NaDone\
formation to the 1D one by showing an $xz$-cut through the simulation
volume along the $y \is 2.0$~Mm dotted line in
Figure~\ref{fig:sim-images} where it passes through the
second-brightest magnetic concentration in panels (b) and (c).  
%The brightest concentration is analyzed in
%Figure~\ref{fig:sim-breakdown-a}.  
The magnetic concentration stands out in panel (b) of
Figure~\ref{fig:sim-3D-1D}, which also illustrates its upward field
spreading, and in panel (d) as a relatively hot deep finger.  The cut
passes through a few more intergranular lanes with fairly strong field
but no vertex concentration, and samples a high cool cloud at $x \is
2.5$~Mm and a low one at $x \is 8.9$~Mm.

The first panel shows the vertically emergent intensity at the nominal
line-center wavelength along the cut, given in equivalent
brightness-temperature units to facilitate comparison with panels
(d)--(j).  The 3D curve has an aureole around the magnetic
concentration and is generally much brighter and smoother than the 1D
curve.  Panel (c) displays the 3D vertical contribution function in
color.  The nominal line-center $\tau_{\nu_0} \is 1$ heights are
overplotted, also in
%RJ ?? tau=1 minimum-per-pixel or nominal line center? Presume nominal
panels (d)--(h) as reference; they do not differ much between 3D and
1D. The line-center formation height varies from $z \approx 0.4$~Mm in
the magnetic concentration to $z \approx1.1$~Mm in the hot rim of the
deep cool cloud that appears as a filamentary feature in the $z \is
0.6$~Mm temperature panel of Figure~\ref{fig:sim-MHD}.  The other
upward spikes in panel (c) sample similar features, all with very wide
contribution functions.  The cool clouds cause $\tau_{\nu_0} \is 1$
dips.

The lower three rows of Figure~\ref{fig:sim-3D-1D} analyze \NaDone\
source function properties, at left for the 1D case and at right for
the 3D case.  Panels (e) and (f) map the profile-averaged
angle-averaged radiation field 
\begin{equation}
  \overline{J} \equiv \frac{1}{4 \pi} \int^\infty_0 \!\!  \int_{0}^{2
    \pi}  \int_{0}^{\pi} I_\nu
  \,\varphi(\nu\!-\!\nu_0) \sin \theta \, \dif\theta \dif\phi \dif\nu,
\end{equation} 
with $I_\nu$ the local specific intensity, $\varphi(\nu\!-\!\nu_0)$
the local \NaDone\ area-normalized extinction profile, $\nu$ the
frequency, $\nu_0$ the nominal line-center rest frequency, and
$\theta$ and $\phi$ specify the viewing angle in spherical
coordinates.  The other panels map the total source function at
$\nu=\nu_0$.  It can be written, dropping frequency indices, as:
\begin{equation}
  S = (1-\varepsilon)\,(1-\eta)\,\overline{J} + \varepsilon\,(1-\eta)\,B(T)
       + \eta\,(1-\varepsilon)\,B^\ast \\
\end{equation}
where $\overline{J}$ represents the \NaDone\ 
photon reservoir, $B(T)$ the thermal reservoir, and $B^\ast$ the
reservoir for roundabout paths (in
particular recombination cascades) that may populate the upper level
and hence produce line photons.  $B^\ast$ is only a formal Planck
function, not necessarily corresponding to the local temperature.  The
coefficient $\varepsilon$ is the thermal line-photon destruction
probability by collisional deexcitation after photoexcitation in the
\NaDone\ transition and through \Hmin\ photoionization as background
continuum process. The profile weighting with the latter contribution
makes the total source function frequency-dependent, even though the
line source function is not for complete redistribution.  The
coefficient $\eta$ is the probability of line-photon conversion via
other ``interlocking'' sodium transitions from the upper level (in
particular photoionization). 
%RR spontaneous + induced

% RR Jefferies unclear as usual; Canfield better
% I included H-min thermalization in eps to make this the total source
% function, so this eps is not just approx C_21/A_21 but combo like RTSA
% 4.93.

% RR oops: I had the S expression wrong in Bruls-I and in SacPeak2005,
% mixing the coefficient definitions.  

% RR oops: complete redistribution case RTSA 4.101 - 4.103 is wrong re
% frequency independence!  OK to define eps as combo 2-level
% destruction and thermal background continuum as in RTSA 4.94 but
% total source function is always frequency-dependent through profile
% weighting, but in RTSA it is called frequency independent below
% 4.103.  It can't be!  RTSA 4.101 and 4.102 must go out and lambda_nu
% must remain monochromatic.  Annotated in Deil RTSA hardcopy.

Panels (e) and (f) are identical in deep layers where $\overline{J}
\approx B$.  Higher up, the two panels appear identical also, in this
case because their structure is dominated by the thermal broadening in
the $\varphi(\nu\!-\!\nu_0)$ profile weighting.  Towards the corona
(note that the red color in panel (d) represents the cutoff at $T \is 10^{4}$~K,
%\krb{(this was only appearance of kK in the paper -- better to stick with plain K)}
% RJ if you want to be fancy, add > in the scale bar
the actual temperatures are higher) the line profile becomes so wide
that $\overline{J}$ approaches the emergent-continuum value with $T_{\rm rad}
\!\approx\! 5100$~K (see the 500~nm panel of Figure~36 of
\cite{1981ApJS...45..635V}, % VALIII
or Figure~3 of \cite{1992A&A...265..237B}). % Bruls++ NaI+KI
In the 1D case this part (orange) is patterned as vertical
search-light beams above hot granules.
% RR correct? Or above steepest T gradients?
They are washed out in the 3D case. Similar wash-out occurs in the
blue-black regime with $T_{\rm rad} < 3600$~K where \NaDone\ is
formed; for example, the cooler areas above $\tau\is1$ in panel (d)
become black in panel (e) but are filled with warmer radiation in
panel (f).  The scattering contribution dominates the source function
in panels (g) and (h) in the blue domain around $z \is 1.0$~Mm, but
higher up the $\varepsilon\,B(T)$ and $\eta\,B^\ast$ contributions mix
in, the first producing the thin bluish-white curve marking the
temperature rise to the corona (where $\varepsilon$ is negligible),
the second producing the slanted greenish features below this curve.
The $\varepsilon\,B(T)$ contribution is the same in 1D and 3D because
it is local, but there is marked difference between 1D and 3D for the
$\eta\,B^\ast$ contribution due to non-local ionizing ultraviolet
radiation.  However, these contributions are too high up to affect the
\NaDone\ brightness, as is demonstrated in panels (i) and (j) by
plotting the source function per vertical column on the relevant part
of the corresponding line-center $\tau_{\nu_0}$ scale.  The magnetic
concentration is the hottest feature crossing the $\tau_{\nu_0}\is1$
line; all other locations sample the blue $\overline{J}$-dominated
scattering regime, of which the columnar structure in 1D explains the
modulation of the corresponding intensity curve in panel (a).  The 3D
curve corresponds to the wash-out of this structure in panel (j).  The
3D source function in panel (h) shows a slanted dynamic fibril
originating from the magnetic concentration, appearing as a
substantial low-reaching green $\eta\,B^\ast$ contribution, but even
this does not affect the last panel because $\tau_{\nu_0} \is 1$
deepens within the concentration.  Thus, this 3D simulation confirms
the 1D conclusion of \citet{1992A&A...265..237B} % Bruls++ NaI+KI
that the source functions of the \NaD\ lines are dominated by
resonance scattering.

\subsection{\NaDone\ formation in specific features}

\begin{figure*}
  \includegraphics[width=\textwidth]{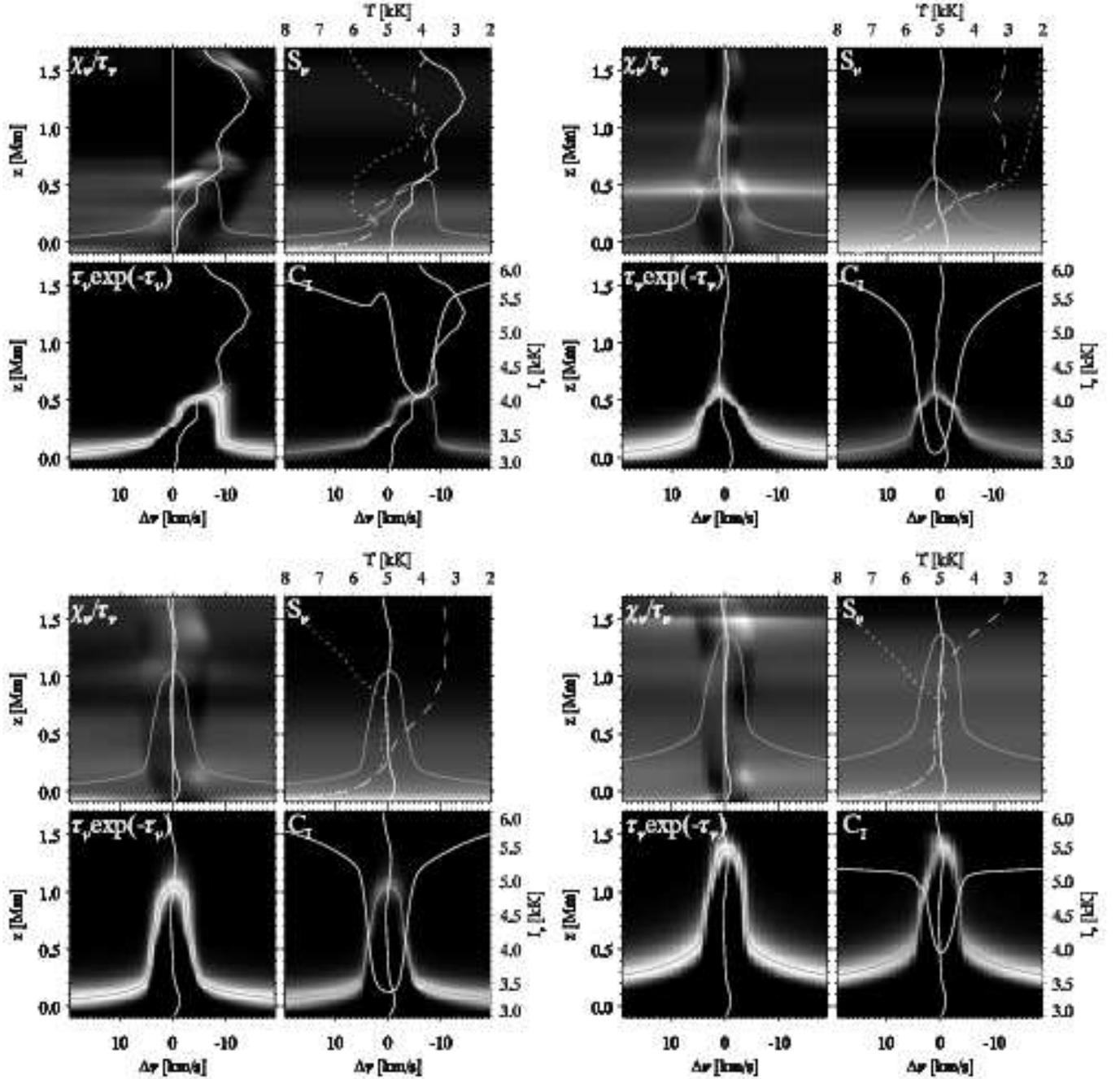}
  \caption{\NaDone\ and \Caline\ formation breakdown for the three
    locations marked in the bottom panel of
    Figure~\ref{fig:sim-images}. The first three quartets are for
    \NaDone, the lower-right one for \Caline.  In each quartet, each greyscale
    image shows the quantity specified in its top-left corner as
    function of frequency from line center (in Dopplershift units) and
    simulation height $z$.  Multiplication of the first three produces
    the intensity contribution function in the fourth panel of each
    quartet. A $\tau_\nu\is1$ curve (grey) and the vertical velocity
    (white) are overplotted in each panel, with a $v_z \is 0$ line in
    the first panel for reference.  The upper-right panel of each
    quartet also contains the Planck function (dotted) and the line
    source function (dashed), in equivalent temperature units
    specified along the top.  The lower-right panel also contains the
    emergent intensity profile, as brightness temperature with the
    scale along the right-hand side. The first quartet is for the
    brightest magnetic concentration at the location marked with a
    plus in the bottom panel of Fig~\ref{fig:sim-images}. The second
    quartet is for the cool cloud marked by the asterisk in the same
    figure. The third quartet is for the hot front marked by the
    rhombus. The final quartet is for the same location but shows the
    formation breakdown for \Caline. }
\label{fig:sim-breakdown}
\end{figure*}

The first quartet of Figure~\ref{fig:sim-breakdown} is a graphical breakdown of \NaDone\
formation for the brightest magnetic concentration marked with a plus
sign in Figure~\ref{fig:sim-images}, using the format developed by
Carlsson \& Stein (1994, 1997)
\nocite{1994chdy.conf...47C} %T Oslo Miniworkshop, H2v grain simulation
\nocite{1997ApJ...481..500C} %C&S CaH&K bright grains
for \CaII\ \HtwoV\ grains.  It rewrites the contribution function to
the vertically emergent intensity $C_I = \rmd I_\nu/\rmd z = S_\nu
\exp(-\tau_\nu)\, \chi_\nu$, with $\chi_\nu$ the total extinction
coefficient, into a product of the three terms $\chi_\nu/\tau_\nu$,
$S_\nu$, and $\tau_\nu \exp(-\tau_\nu)$ that are plotted as greyscale
quantities in the first three panels with $C_I$ as their product in
the fourth one.  The three terms in this decomposition demonstrate
that a large contribution to intensity is made at those positions that
have a large opacity at low optical depth, with a high source function
around optical depth unity. The line source function in the second
panel is frequency-independent because of the assumption of complete
redistribution. Strictly speaking one should plot the total source
function here. However, to increase readability we avoid the cluttered
modulation of the frequency-dependent total source function and plot
the line source function instead. This is justified because the line
source function becomes equal to the continuum source function in the
deep-formed LTE wings, and dominates the total source function in the
scattering line core.

The emergent \NaDone\ profile in the last panel of the first quartet is very asymmetric. It
shows a bright hump at nominal line center that causes the
extraordinary brightening of this feature in the second panel of
Figure~\ref{fig:sim-images}. The actual line core is redshifted over
7~km~s$^{-1}$. The formation mechanism of this profile is rather
similar to the formation of internetwork \CaII\ \HtwoV\ grains. A
large (10~\kms) downdraft in high layers moves the core opacity
redward.  Its inward integration at the profile minimum reaches $\tau
\is 1$ at $z \is 0.5$~Mm, where the source function is dominated by
scattering and has uncoupled from the Planck function. The profile
minimum is much darker than the emergent intensity at nominal line
center. However, it still stands out in the Dopplershift-compensated
panels of Figure~\ref{fig:sim-images}, evidencing a relatively high
source function notwithstanding the signature loss by
scattering. Figure~\ref{fig:sim-images} suggests that the same occurs
in other magnetic concentrations.

Several circumstances conspire to produce the bright hump in the
profile at nominal line center. Because of the gradient in the
downflow velocity there is a location with high $\chi_\nu/\tau_\nu$ at
$\Delta \nu \is 0$ and $z \is 0.3$~Mm (upper-left panel). This
location happens to lie close to optical depth unity at that frequency
(lower-left panel). Most importantly, the source function is not yet
completely set by scattering, and partially follows the increase of
the Planck function above $z \is 0.1$~Mm (dotted line, upper-right
panel). This gives rise to the local maximum in the source function in
the upper-right panel of the first quartet. All effects combined yield a high contribution
function (lower-right panel) and thus a peak in the profile. There is
no corresponding peak at the red side of the profile because
$\chi_\nu/\tau_\nu$ (upper-left panel) is low where $\tau_\nu \is 1$
coincides with the location of the source function maximum because of
the velocity gradient in the atmosphere.

The second quartet of Figure~\ref{fig:sim-breakdown} shows a similar breakdown at the
asterisk in the cool cloud in the bottom panel of
Figure~\ref{fig:sim-images}.  In this case the profile asymmetry is
minor, there are only a small downdraft below $z \is 0.3$~Mm and a
small updraft above this height.  The Planck function (dotted) drops
steeply to 2400~K at $z \is 1.0$~Mm. The source function happens to
follow it far out, but it actually uncouples already at $z \is 0.2$~Mm
where scattering sets in.  It levels out at a radiation temperature
3100~K near $z \is 0.6$~Mm where $\tau_{\nu_0} \is 1$.  The resulting
line core is very dark.
 
The third quartet of Figure~\ref{fig:sim-breakdown} shows another breakdown for the
slender hot front marked with a rhombus in the bottom panel of
Figure~\ref{fig:sim-images}, where \NaDone\ has maximum formation
height with $\tau_{\nu_0} \is 1$ at $z \is 1.1$~Mm.  The Planck
function has an extended minimum for $T \approx 5000$~K and then rises
steeply with height, through 6000~K at $z \is 1.1$~Mm, but this is not
followed by the source function which uncouples much deeper and has a
radiation temperature of 3400~K at $z \is 1.1$~Mm.  Thus, although
\NaDone\ is formed very high at this location, it does not show the
relatively high temperature there. Note the large difference in
line-core formation height between the cool cloud of
the second quartet and the hot front of
third quartet. This is caused by the strong
dependence of the \NaDone\ opacity on the electron density. The cool
cloud has a low temperature, hence a low LTE electron density and thus
a low opacity due to relative lack of photorecombination (see Sec \ref{sec:emergent}). The hot
front has a relatively high temperature throughout the whole column,
so that the relatively high opacity results in a large formation
height.

For comparison, the fourth quartet shows the
corresponding \Caline\ formation breakdown at the location of the hot front, confirming that this line
is formed higher and has better coupling to the temperature: this
location is quite bright in \Caline\ (Figures~3 and 4 of
Paper~I).

% The latter does not arise from shock dynamics along this column
% since the vertical velocity is nearly zero at all heights.  It is
% probably produced by slanted shocks, as suggested by the patterning
% along the cut through this location in Figure~4 of Paper~I.

\section{Discussion} \label{sec:discussion}

\subsection{LTE ionization versus non-equilibrium ionization}

The MHD simulation is quite realistic in the photosphere but arguably
less so higher up.  One obvious imperfection is the assumption of
LTE Saha-Boltzmann partitioning for hydrogen and the electron-donor
metals.  These are largely neutral in the simulation's cool clouds,
but sodium gets largely ionized there in our subsequent NLTE
statistical-equilibrium modeling for that element because its
ionization is dominated by hot photospheric radiation.  The same would
be the case for the other electron donors in such modeling.  Worse,
\citet{2002ApJ...572..626C} % Carlsson+Stein dynamic H ionization
and \citet{2007A&A...473..625L} %C Leenaarts hion2
have shown that for hydrogen the assumption of statistical equilibrium
fails also, requiring the evaluation of non-equilibrium
ionization--recombination balancing in which the past history of a
given gas parcel affects its state, in particular suffering slow
recombination in the cool aftermath of a fast-ionizing hot shock
passage.  Similar slowness may affect other electron-donor elements,
although hydrogen is likely the worst.  Therefore, the low electron
densities in the cool clouds in Figure~\ref{fig:sim-MHD}, with their
large effect on the \NaDone\ $\tau =1$ height in the bottom panel of
Figure~\ref{fig:sim-images}, are likely artifacts.  The actual
electron density may remain as high as $N_\rme/N_\rmH \approx 10^{-2}$
in shocked gas, causing up to $10^4$ increase in \Hmin\ opacity and
therefore increased radiative heating above the photosphere wherever
the \Hmin\ bound-free continuum has $\overline{J_\nu} > B_\nu$.  So
perhaps the cool clouds will not become so cool in the case of
non-equilibrium ionization.  On the other hand, slow balancing affects
the energy budget intrinsically by not instantaneously converting
thermal energy into ionization or releasing it in recombination,
%RR S-B assumes collisional recombination = thermal energy release
resulting in larger temperature contrast across shocks
(\cite{1992ApJ...397L..59C}) %Carlsson+Stein 1D(t) shocks
and longer-lived cooler post-shock temperatures
(\cite{2007A&A...473..625L}). % Leenaarts++ hion2
It is hard to predict what cool clouds future simulations with
improved ionization treatment will produce, but probably these will have
larger electron density than the cool clouds here, and possibly larger
\NaDone\ opacity.

\subsection{Chromosphere or clapotisphere}

Since we use only a single snapshot of the simulation here, we cannot
identify the nature of its fine structure in higher layers; that would
require analysis of its temporal evolution.  Nevertheless, comparison
with the 2D simulation of
\citet{2007A&A...473..625L} %C Leenaarts hion2
and the 3D simulation of
\citet{2009ApJ...701.1569M} % Martinez-Sykora et al: spicule simulations
suggests that many of the fine-scale features in
Figures~\ref{fig:sim-MHD}--\ref{fig:sim-3D-1D} are due to shock
dynamics, with slanted wave-guiding along field lines in the
neighborhood of the magnetic concentrations producing short dynamic
fibrils (see 
\cite{2006ApJ...647L..73H}; % Hansteen++ dynamic fibrils
\cite{2007ApJ...655..624D}; % DePontieu++ dynamic fibrils
Paper~I).  The scene in the rightmost column of
Figure~\ref{fig:sim-MHD} consists of slender hot shock fronts and cool
expansion clouds.  There appear to be no long fibrils in the
simulation that would be comparable to those typically observed near
network in \Halpha\ and in \Caline\ (Figure~\ref{fig:obs-filter}).
This is consistent with the observed \NaDone\ profile minimum
intensity image shown in Fig.~\ref{fig:obs-filter} which also shows
almost no evidence of fibrils in this quiet region.

The simulated \Caline\ line-center image from the same snapshot in
Figure~3 of Paper~I does not contain such long fibrils either, and
shows larger disparity with the observed line-center scene than occurs
here between panel (e) of Figure~\ref{fig:obs-center} and panel (d) of
Figure~\ref{fig:sim-images}.  We therefore conclude that the
simulation does not contain the dense carpets of extended fibrils that
constitute the magnetically dominated chromosphere, but that it models
what we call the clapotisphere quite well. The simulation is a fair
description of what is observed in \NaDone\ but is deficient
in \Halpha-type fibrils in \Caline.  Such deficiency doesn't really
matter for \NaDone\ because its intensity is determined well below the
clapotispheric regime anyhow: although its $\tau_{\nu_0} \is 1$ height
in the third quartet of Figure~\ref{fig:sim-breakdown} is clapotispheric, the source
function there is dominated by $\overline J$ photons originating much
deeper.  Only the core Dopplershift is set at clapotispheric heights,
but only outside the transparent cool clouds.

\subsection{Magnetic concentrations in \NaDone\ and \Caline}

The magnetic concentrations appear bright in both the observations and
the simulation, even after compensation for the Dopplershift imposed
by the large downdrafts that tend to occur at their locations, also
both in the observations and the simulation.  In the simulation this
brightness is due to higher-than-elsewhere temperatures in the upper
photosphere ($z \is 0.3$~Mm temperature panel of
Figure~\ref{fig:sim-MHD}).  

Near and within magnetic concentrations, field-guided shocks tend to
appear deeper than elsewhere, as was already shown in the classical 2D
simulation of
\citet{1998ApJ...495..468S} % Steiner++, Freiburg fluxsheet
and confirmed later by the dynamic fibril studies of
\citet{2006ApJ...647L..73H} % Hansteen++ dynamic fibrils
and 
\citet{2007ApJ...655..624D}. % DePontieu++ dynamic fibrils
The strong narrow downdrafts in most magnetic concentrations (bottom
row of Figure~\ref{fig:sim-MHD}) may be post-shock downdrafts.  As
diagnosed in the first quartet of Figure~\ref{fig:sim-breakdown}, they cause
magnetic-concentration brightening with a similar formation mechanism
as the asymmetric \CaII\ \HK\ violet-wing brightening in non-magnetic
\HtwoV\ and \KtwoV\ grains. These sample the sub-canopy shocks in the
internetwork at $z \approx 1$~Mm.  
\NaDone\ is formed much lower and samples only the deep-sited shocks
in and near magnetic concentrations.  The formation of \Caline\ is
intermediate; the large brightness of the magnetic
concentration in Figure~4 of Paper~I is similarly
formed through higher-up downdraft.

The breakdown for the magnetic concentration in Figure~\ref{fig:sim-breakdown} shows a raised
temperature (Planck function in the second panel) over $z \is
0.3-0.8$~Mm; it is the hottest location in the $z \is 0.3$~Mm
temperature panel of Figure~\ref{fig:sim-MHD}.  The other magnetic
concentrations have similarly raised temperatures at this height.  The
corresponding source function hump at $z \is 0.3$~Mm, which produces
the blue-wing peak at nominal line center, lies within the upper
photosphere outside, in the reversed granulation regime
(Figure~\ref{fig:sim-MHD}).

The vertical velocity curve in the first quartet Figure~\ref{fig:sim-breakdown}
(overplotted in all panels) displays a series of steps suggesting
shock passages.  If so, these are less well defined, with less obvious
response in temperature, than those in Figures~5--7 of
\citet{1997ApJ...481..500C} %C C+S H2v grain simulation
from their non-magnetic $\HtwoV$ grain simulation with the 1D RADYN
code.  That code used an adaptive mesh and had much better vertical
resolution than the fixed mesh of OSC.  Also, in 3D shocks are likely
not to run up straight even within magnetic concentrations.  Analysis
of the full time-dependent simulation will tell whether this velocity
pattern indeed stems from shocks, and may establish whether shock
heating matters in setting this deep-seated temperature increase in
magnetic concentrations, as suggested by the high correspondence
between excess temperature and excess downdraft in magnetic
concentrations in the second column of Figure~\ref{fig:sim-MHD}.

\subsection{Comparison between observation and simulation}

The observed and simulated scenes in panel (e) of
Figure~\ref{fig:obs-center} and panel (d) of
Figure~\ref{fig:sim-images} show similar morphology. Within the
simulation, \NaDone\ is primarily a magnetic-concentration mapper in
its brightness distribution, sampling these in the upper photosphere
and including wide scattered-light aureoles. Elsewhere, the
line-center intensities are fully dominated by scattering, producing
very dark source function sampling with much wash-out of thermal
structure. Since these formation characteristics are primarily
photospheric, where we believe our simulation to be realistic, we
suggest that the same apply to the Sun.

% RK ?? here add Kevin contrast-across-profile analysis and dicussion?
% I don't know what to make of it.  We now concentrate on line center
% formation.  Maybe comparison core Dopplershift contrasts is of most
% interest but that answer is already clear - compare obs-(f) sim-(f)
% scenes and scales.  Very disparate.  Written up below.

% RR perhaps interesting to produce Kevin contrast-across-profile
% plots also for 8542.  And to have NaD+8542 from CRISP and compare.

In the simulation only the core Dopplershifts, imposed at the last
scattering, are likely to sample clapotispheric heights, and that only
outside the cool clouds where sodium is fully ionized.  The
Dopplershift map in panel (f) of Figure~\ref{fig:obs-center} has
similar slender magnetic-concentration downdrafts as in the companion
panel (f) of Figure~\ref{fig:sim-images} but less other fine
structure, in particular missing out on the considerable downflows in
cool clouds and the sharp, slender upflow features from hot shocks
that the simulation presents.  These differences may be attributable
to lack of resolution in the observations and/or overestimation of
clapotispheric dynamics in the simulation.  At this stage, we can't
tell.

% RR The 8542 paper has CRISP giving more similar fine structure at
% knee wavs.  No core Doppler maps there, would be interesting.  We
% could add them here but that asks for z=1200 km column in
% fig:sim-MHD etc.  And a bit besides the core messsage = NaD
% formation.

\section{Conclusions} \label{sec:conclusions}

Based on the analysis of our simulation we conclude that the brightest
quiet-Sun features observed in the core of \NaDone\ are also the
deepest: magnetic concentrations sampled at $z\approx 0.3-0.5$~Mm.
The lower height holds when one does not compensate for Dopplershifts
in the line-center sampling and a core shift causes additional
brightening.  This happens in particular when the magnetic
concentrations harbor higher-layer downdrafts producing blue-wing
peaks through the same line-formation mechanism that causes
non-magnetic internetwork \HtwoV\ and \KtwoV\ grains in \CaII\ \HK.
Such downdrafts occur in most magnetic concentrations in the
simulation and seem to occur regularly also in the observations,
although even our high-resolution IBIS imaging barely resolves them.

The next brightest component in quiet-Sun \NaDone\ scenes are
relatively bright aureoles surrounding the magnetic concentrations,
due to resonance scattering of \NaD\ radiation from the magnetic
concentrations.  This is very much a 3D phenomenon requiring 3D NLTE
modeling, as demonstrated in Figure~\ref{fig:sim-3D-1D}.

Elsewhere, quiet-Sun \NaDone\ scenes are dark even though at certain
locations the formation height may be located much higher up, around
$z \is 1$~Mm within hot clapotispheric shock fronts.  Their high
temperature has no response in the \NaDone\ source function due to the
strong resonance scattering in this line, but it does for the \Caline\
line which combines higher formation with better temperature
sensitivity.

The upshot is that in quiet-Sun areas the \NaDone\ core displays
primarily an upper-photosphere rendering of magnetic concentrations on the
solar surface, with roughly the same surface patterning as in unsigned
magnetograms or in proxy diagnostics such as G-band and \Halpha-wing
bright points, but contaminated with brightening from downdrafts and
3D-scattered aureoles.  This similarity explains the correlation
between photospheric magnetic flux density and \NaDone\ core
brightness reported by
\citet{2000A&A...357.1093C}, % Cauzzi++ NaD vs HK
and it makes \NaDone\ just another photospheric-field indicator in
irradiance monitoring.  It is not a chromosphere diagnostic, not even
a useful clapotisphere one.  Only usage of information imposed at the
last scattering (Dopplershift, polarization) seems of interest, but
that suffers from the large variation in the height of the $\tau \is
1$ surface at line center.

The magnetic concentrations in the simulation are hotter at the $z \is
0.3-0.5$~Mm \NaDone\ line-center sampling height than the outside
upper photosphere with reversed granulation. Further analysis of the
full time-dependent simulation may clarify how this heating comes
about; \NaDone\ may be useful to constrain the mechanism.

%%%%%%%%%%%%%%%%%%%%%%%%%%%%%%%%%%%%%%%%%%%%%%%%%%%%%%%%%% ACKNOWLEDGEMENTS
\begin{acknowledgements}
  JL acknowledges financial support by the European Commission through
  the SOLAIRE Network (MTRN-CT-2006-035484).  RJR acknowledges travel
  support from the USO-SP International School for Solar Physics
  funded by the European Commission (MEST-CT-2005-020395). This
  research was supported by the Research Council of Norway through
  grant 170935/V30 and through grants of computing time from the
  Programme for Supercomputing.  The IBIS observing relied on DST
  observers M.~Bradford, J.~Elrod, and D.~Gilliam. 
  We also thank Ryoko Ishikawa for her collaboration in obtaining the
  Hinode and IBIS observations.
  IBIS was
  constructed by INAF/OAA with contributions from the University of
  Florence, the University of Rome, MIUR, and MAE, and is operated
  with support of the US National Solar Observatory (NSO).  The NSO is
  operated by the Association of Universities for Research in
  Astronomy, Inc., under cooperative agreement with the US National
  Science Foundation. Hinode is a Japanese mission developed and
  launched by ISAS/JAXA, with NAOJ as domestic partner and NASA and
  STFC (UK) as international partners. It is operated by these
  agencies in co-operation with ESA and NSC (Norway). The SOT/SP Level
  1D data was generously provided by Bruce Lites. This study benefited
  from fruitful discussions at and supported by the International
  Space Sciences Institute at Bern, Switzerland, and from NASA's
  Astrophysics Data System.
\end{acknowledgements}

%%%%%%%%%%%%%%%%%%%%%%%%%%%%%%%%%%%%%%%%%%%%%%%%%%%%%%%%%%%%%%%% REFERENCES
\bibliographystyle{aa}
%\bibliography{aajour,/tmp/rjrfiles,/tmp/adsfiles,cauzzi}
\bibliography{nordlund,%
cavallini,%
leenaarts,%
cheung,%
vitas,%
keller,%
reardon,%
sanchez-almeida,%
carlsson,%
cauzzi,%
rimmele,%
wedemeyer,%
hansteen,%
martinez-sykora%
,carlson,%
neckel,%
bruls,%
fontenla,%
spruit,%
solanki,%
berger,%
rouppe,%
lites,%
rutten,%
vernazza,%
depontieu,%
steiner,%
uitenbroek,%
kneer,%
skartlien,%
botnen,%
gustafsson}

%RR oops, I got newest natbib.sty but it doesn't work
%RR @R improve ADS conference items as usual
%RR @R add Paper~I

%%%%%%%%%%%%%%%%%%%%%%%%%%%%%%%%%%%%%%%%%%%%%%%%%%%%%%%%%%%%%%%%%%%%%%%%%%%%
%%\input{nadfigs} %RJ @J insert here for submission with aastex.cls

%%%%%%%%%%%%%%%%%%%%%%%%%%%%%%%%%%%%%%%%%%%%%%%%%%%%%%%%%%%%%%%%%%%%%%% END
\end{document}

%% file: rrmacros2e-AA.tex
\newcount\longrefs
\def\aap{\ifnum\longrefs=1 {Astron.\ Astrophys.}\else 
                           {A\hbox{\rm \&}A}\fi}
\def\aapr{\ifnum\longrefs=1 {Astron.\ Astrophys.\ Rev.}\else 
                            {A\hbox{\rm \&}AR}\fi}
\def\aaps{\ifnum\longrefs=1 {Astron.\ Astrophys.\ Suppl.}\else 
                            {A\hbox{\rm \&}A Suppl.}\fi}
\def\aj{\ifnum\longrefs=1 {Astron.\ J.}\else 
                          {AJ}\fi} 
\def\ao{\ifnum\longrefs=1 {Applied Optics}\else 
                           {Appl.\ Opt.}\fi} 
\def\aspcs{\ifnum\longrefs=1 {Astron.\ Soc.\ Pacific Conf. Series}\else 
                           {ASP Conf.\ Ser.}\fi} 
\def\apj{\ifnum\longrefs=1 {Astrophys.\ J.}\else 
                           {ApJ}\fi} 
\def\apjl{\ifnum\longrefs=1 {Astrophys.\ J. Lett.}\else 
                            {ApJ}\fi} 
\def\aplett{\ifnum\longrefs=1 {Astrophys.\ J. Lett.}\else 
                            {ApJ}\fi} 
\def\apjs{\ifnum\longrefs=1 {Astrophys.\ J. Suppl.}\else 
                            {ApJS}\fi}
\def\apss{\ifnum\longrefs=1 {Astrophys.\ and Space Science}\else 
                            {Astrophys.\ Space Sci.}\fi}
\def\araa{\ifnum\longrefs=1 {Ann.\ Rev.\ Astron.\ Astrophys.}\else 
                            {ARA\hbox{\rm \&}A}\fi}
\def\azh{\ifnum\longrefs=1 {Astronomicheskii Zhurnal}\else 
                            {Astron.\ Zhur.}\fi}
\def\baas{\ifnum\longrefs=1 {Bull.\ Am.\ Astron.\ Soc.}\else 
                            {BAAS}\fi}
\def\bain{\ifnum\longrefs=1 {Bull.\ Astronom.\ Institutes Netherlands}\else
                            {Bull.\ Astr.\ Inst.\ Neth.}\fi}
\def\gca{\ifnum\longrefs=1 {Geochim.\ Cosmochim.\ Acta}\else 
                           {Geochim.\ Cosmochim.\ Acta}\fi}
\def\grl{\ifnum\longrefs=1 {Geophys.\ Res.\ Lett.}\else 
                           {Geoph.\ Res.\ Lett.}\fi}
\def\iaucirc{\ifnum\longrefs=1 {IAU Circulars}\else 
                          {IAU Circ.}\fi}
\def\ip{\ifnum\longrefs=1 {in press}\else 
                          {in press}\fi}
\def\jgr{\ifnum\longrefs=1 {J.\ Geophys.\ Res.}\else 
                           {J.\ Geophys.\ Res.}\fi}  
\def\jrasc{\ifnum\longrefs=1 {J.\ Royal Astron.\ Soc.\ Canada}\else 
                           {JRAS Can.}\fi}  
\def\mnras{\ifnum\longrefs=1 {Mon.\ Not.\ Roy.\ Astron.\ Soc.}\else 
                             {MNRAS}\fi} 
\def\nat{\ifnum\longrefs=1 {Nature}\else 
                           {Nat}\fi}
\def\pasj{\ifnum\longrefs=1 {Pub.\ Astron.\ Soc.\ Japan}\else 
                            {PASJ}\fi} 
\def\pasp{\ifnum\longrefs=1 {Pub.\ Astron.\ Soc.\ Pacific}\else 
                            {PASP}\fi} 
\def\physscr{\ifnum\longrefs=1 {Physica Scripta}\else 
                            {Phys.\ Scrip.}\fi} 
\def\planss{\ifnum\longrefs=1 {Planetary \& Space Science}\else 
                            {Plan. \& Space Sci.}\fi} 
\def\procspie{\ifnum\longrefs=1 {Proc.\ SPIE}\else 
                            {Proc.\ SPIE}\fi} 
\def\qjras{\ifnum\longrefs=1 {Quarterly J.\ Royal Astron.\ Soc.}\else 
                            {QJRAS}\fi} 
\def\sa{\ifnum\longrefs=1 {Soviet Astron..}\else 
                               {Sov.\ Astron.}\fi}
\def\skytel{\ifnum\longrefs=1 {Sky \& Telescope}\else 
                            {Sky \& Tel.}\fi} 
\def\solphys{\ifnum\longrefs=1 {Solar Phys.}\else 
                               {Sol.\ Phys.}\fi}
\def\ssr{\ifnum\longrefs=1 {Space Science Rev.}\else 
                               {Space\ Sci.\ Rev.}\fi}

\def\rmit#1{{\it #1}}              %% italics (RR style, Kluwer)
\def\etal{\rmit{et al.}}           %% use \etal\ for space behind it        
           
\def\ie{\rmit{i.e.,}}              %% , required (Webster 1681)
\def\eg{\rmit{e.g.,}}              %% , required (Webster 1681)
\def\cf{cf.}                       %% no Latin, always Roman (Webster 1686)

\def\specchar#1{\uppercase{#1}}    %% to be redefined for A&A, small caps
\def\CaII{\mbox{Ca\,\specchar{ii}}}

\def\FeI{\mbox{Fe\,\specchar{i}}}

\def\Hmin{\hbox{\rmH$^{^{_-}}\!$}}      %% H^min, very elegant
\def\NaI{\mbox{Na\,\specchar{i}}}
\def\NaII{\mbox{Na\,\specchar{ii}}}

\def\Halpha{\mbox{H\hspace{0.1ex}$\alpha$}} %% \Halpha\ for space behind it

\def\NaD{\mbox{Na\,\specchar{i}\,D}}    %% use \NaD\ for space behind it
\def\NaDone{\mbox{Na\,\specchar{i}\,\,D$_1$}}
\def\NaDtwo{\mbox{Na\,\specchar{i}\,\,D$_2$}}

\def\HK{\mbox{H\,\&\,K}}
\def\KtwoV{\mbox{K$_{2V}$}}

\def\HtwoV{\mbox{H$_{2V}$}}

\def\rmd{{\rm d}}  
\def\rme{{\rm e}}

 \def\rmH{{\rm H}}

\def\deg{\hbox{$^\circ$}}       %% \def for overwriting, \box for math

\def\kms{\hbox{km$\;$s$^{-1}$}}

\def\is{\!=\!}                             %% = in text for tighter spacing
\def\dif{\: {\rm d}}                       %% differential d with space
\def\={\hbox{$\!=\!$}}                     %% less space around =